# Evidence for biological Lévy flights stands

Denis Boyer[1], Octavio Miramontes[1] & Gabriel Ramos-Fernández[2]

[1]*Departamento de Sistemas Complejos, Instituto de Física, Universidad Nacional Autónoma de México, Apartado Postal 20-364, 01000 México DF, México*

[2]*Instituto Politécnico Nacional, CIIDIR Unidad Oaxaca, Calle Hornos 1003, Santa Cruz Xoxocotlán, 71230 Oaxaca, Mexico*

**Edwards *et al.*[1] revisited well-known studies reporting power-laws in the frequency distribution of flight duration of wandering albatrosses[2,3], and concluded that no Lévy process[4,5] could model recent observations with higher resolution. Here we show that this re-analysis[1] suffers from a conceptual misunderstanding, and that the new albatross data remain consistent with a biological Lévy flight.**

Edwards *et al*. focus the interpretation of their results on the tails of the frequency distributions of foraging movements (or their rank/frequency plots, RFP). The exponential law, not the power-law, dominates the 2004 high resolution albatross data at large scales, indicating a Poisson process[1]. However, Edwards *et al*. have underestimated the crucial fact that all foraging processes are subject to finite size constraints. Exponential tails are actually an essential outcome (although overlooked so far) of Lévy random search theory[3]: they stem from target detection, a foraging issue by excellence. A forager with a perception radius $r$ and deciding to move at constant velocity $v$ during a time $x$ on a plane containing randomly distributed targets in number density $\rho$ has a probability $e^{-x/\tau}$ of not finding any target, with $\tau = (2vr\rho)^{-1}$. We derive the *actual* step duration distribution for the model of ref. 3:

$$P(x) = P_0(x)\, e^{-x/\tau} + \tau^{-1}\, e^{-x/\tau} \int\limits_{x}^{\infty} du P_0(u) \qquad (1)$$

with $P_0(x) = C x^{-\mu}$ ($\mu > 1$) being the *choice* distribution, and $C$ a normalization constant. The first term in (1) is the probability of making a trip of duration $x$ and not finding anything, the second the probability of finding a target after a time $x$ (implying that the chosen time $u$ is $> x$). $P$ takes two limiting forms: *(i)* $P(x) \approx C x^{-\mu}$ if $x \ll \tau(\mu - 1)$, a wide interval only if resources are scarce; *(ii)* $P(x) \sim x^{1-\mu}\, e^{-x/\tau} \neq P_0(x)$ if $x \gg \tau(\mu - 1)$. Formula (1) has sounder biological and physical grounds than the *ad hoc* gamma function introduced in ref. 1 to fit the albatross data. It should also be considered as an alternative to the power-law distribution, obtained when $\tau \to \infty$, for testing biological Lévy flights. With $\mu = 1.18$ and $\tau = 1.89$ hours, Eq. (1) describes the albatross data very well (Fig.1). The data is not consistent with a pure Poisson process, $P(x) = \tau^{-1} e^{-x/\tau}$, which is recovered in Eq. (1) when $\mu \to 1$ (Fig. 1). Although $\mu$ differs from the optimal value 2 for immobile targets[3], it remains larger than unity: the flying times can still be interpreted as drawn from a genuine, normalizable power-law distribution, contrary to what is concluded in ref. 1 ($\mu = 0.69$).

Furthermore, the lack of straight lines in a log-log RFP (as in our fig. 1, or in figs. 1 and 3c of ref. 1) is not conclusive evidence for the absence of a power-law pattern. Truncated-power-law frequency distributions do not produce RFP with straight lines in log-log, even at small scales, unless their scaling regime spans over at least three decades. This is practically never the case in foraging data.

Likelihood and goodness-of-fit tests are useful methods to rule out hypotheses, or to conclude that several models (e.g. Poisson, Lévy) are equally likely to describe a given data set of small size, as illustrated by Edwards *et al.* in the case of bumble-bees[1]. But in order to improve the understanding of foraging processes, these tests should be applied, whenever possible, to analytical or numerical results of foraging models, rather than to *a priori* given mathematical functions with limited biological content.

1. Edwards, A.M. *et al*. *Nature* **449,** 1044-1048 (2007).

2. Viswanathan, G.M. et al. *Nature* **381**, 413–415 (1996).

3. Viswanathan, G.M. et al. *Nature* **401**, 911–914 (1999).

4. Shlesinger, M.F. & Klafter, J. In *On Growth and Form*, (eds. Stanley H.E. & Ostrowski N.) 279–283 (Martinus Nijhoff, Amsterdam, 1985).

5. Klafter, J. *et al.* In *Biological Motion*, (eds. Hoffmann, G. & Alt, W.) 281–296 (Springer, Heidelberg, 1989).

6. Sokal, R.R. & Rohlf, F.J. *Biometry: the principles and practices of statistics in biological research*, 3$^{rd}$ edition (Freeman, New York, 1994).

**Figure 1**. Cumulative distribution of the duration of the 2004 albatross flights[1] (*x*, open circles), compared with the cumulative of the probability distribution function predicted by Lévy search theory, Eq. (1), with $\mu$=1.18 and $\tau$=1.89h (solid line). Since only the flights larger than 30s were considered in the field study[1], we have taken the power-law choice distribution $P_0(x) \neq 0$ if *x*>30s and 0 otherwise. With this assumption the forager can still perform flights that have *x*<30s in the model (1); nevertheless, only those with *x*>30s are represented, as in the original albatross data. These data were obtained by digitalizing the histogram of Ref. 1. A log-likelihood ratio test of goodness-of-fit (G-test[6]) was performed with the parameter values above for $\mu$ and $\tau$ by using a Monte Carlo (MC) procedure generating $10^4$ independent data sets. We obtain *P*=0.21, meaning that 21% of the independent data sets drawn from Eq. (1) are further away from this distribution than the albatross data (G=53.6, degrees of freedom=47). The two dotted curves are from Eq. (1) with $\mu$=1.01 ($\tau$=1.89h and 1h): at this exponent value, the distribution *P(x)* is nearly Poisson and cannot fit the data for any $\tau$ (*P*<$10^{-4}$). The discrete nature of the albatross data (see Ref. 1, Supplementary Methods 1) was taken into account in the MC calculations and for plotting the theoretical curves.

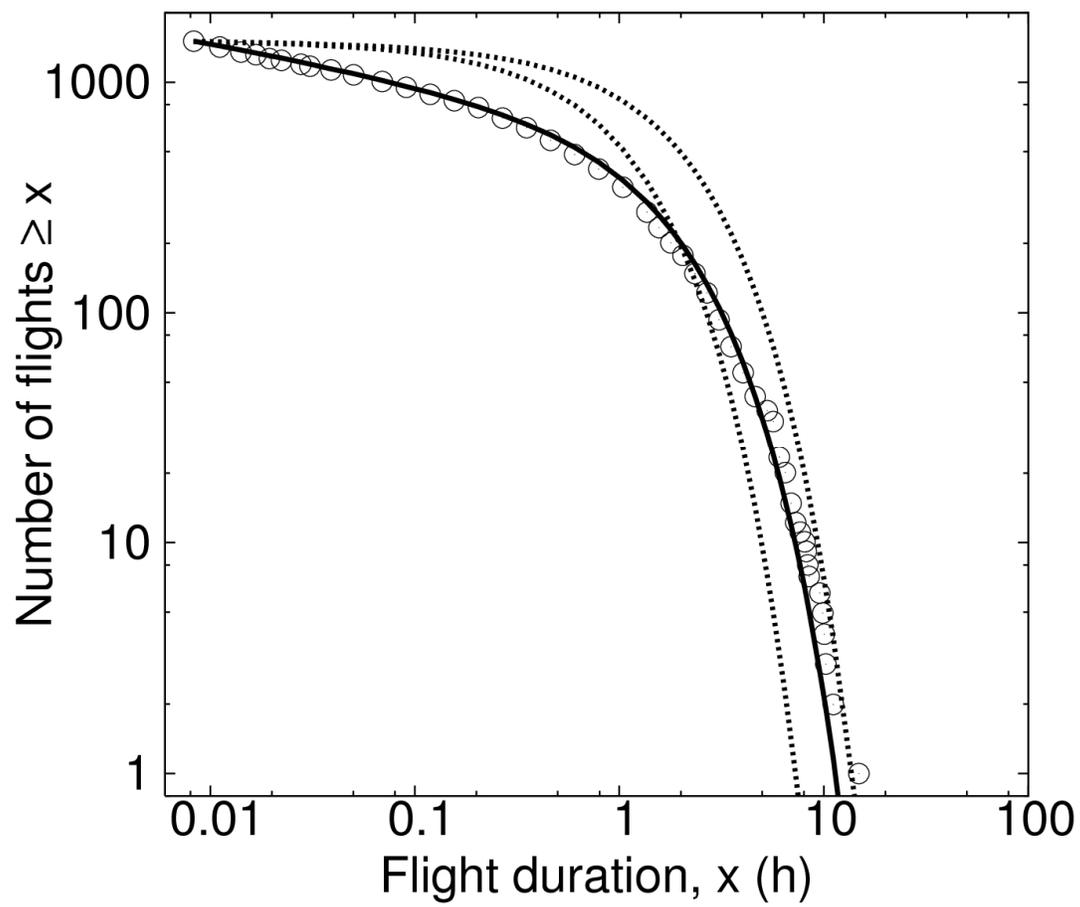